\documentclass[12pt]{article}
\usepackage{sw20lart}

\input tcilatex

\begin{document}

\textbf{DYNAMICS\ OF\ DISTRIBUTED\ SOURCES}\vspace{0.3in}

E.A. Novikov\vspace{0.2in}

Institute for Nonlinear Science, University of California / San Diego

La Jolla, CA 92093-0402\vspace{0.3in}

Dynamics of distributed sources is described by nonlinear partial
differential equations. Lagrangian analytical solutions of these (and
associated) equations are obtained and discussed in the context of the
Lagrangian modeling - from the Lagrangian Invariants to dynamics. Possible
applications of distributed sources and sinks to geophysical fluid dynamics
and to gravitation are indicated.\vspace{0.3in}

Singular vortices, sources (sinks) and vortex-sinks are well known models in
fluid dynamics (including geophysical fluid dynamics), magnetized plasma,
superfluidity and superconductivity (see, for example, Refs. 1-3 and
references therein). Dynamics of distributed vortices is well developed
classical area of research. However, to our knowledge, dynamics of
distributed sources has not been considered before. Structure and intensity
of distributed sources can change in time and, in this sense, they are more
rich and flexible models for various media than singular sources with fixed
intensity. In this letter we show that distributed sources have a nontrivial
dynamics, which in some cases can be described analytically in Lagrangian
variables.

Local intensity of distributed sources (sinks) in moving continuous media is
characterized by the divergency of the velocity field $\alpha (t,\mathbf{x}%
)=\partial v_i(t,\mathbf{x})/\partial x_i$, where summation over the
repeated subscripts is assumed from $1$ to the dimension of the space $%
s=1,2,3$. To get an equation for $\alpha $ , we consider superposition of
localized sources, which move each other with induced velocity field$%
^{[2,3]} $. Thus, the equation can be written in form: 
\begin{equation}
\frac{d\alpha }{dt}\equiv \frac{\partial \alpha }{\partial t}+v_k\frac{%
\partial \alpha }{\partial x_k}=f  \tag{1}
\end{equation}
Here $f$ may include diffusion and some other effects (see below). We start
with consideration of ``free'' sources with $f=0$. In this case $\alpha $ is
the Lagrangian Invariant ($LI$), which is conserved along trajectories. For
the potential motion with velocity $v_i=\partial \varphi /\partial x_i$,
from (1) we get: 
\begin{equation}
\frac{d\alpha }{dt}\equiv \frac{\partial \Delta \varphi }{\partial t}+\frac{%
\partial \varphi }{\partial x_k}\frac{\partial \Delta \varphi }{\partial x_k}%
=0,\alpha =\Delta \varphi ,\Delta \equiv \frac{\partial ^2}{\partial x_i^2} 
\tag{2}
\end{equation}

The moving media, described by equation (2), can be interpreted as the
barotropic (or incompressible) fluid: $dv_i/dt=-\partial w/\partial x_i$
with $w$ determined by the equation $\Delta w=-\partial v_i/\partial
x_k\partial v_k/\partial x_i=-(\partial ^2\varphi /\partial x_i\partial
x_k)^2$. The equation for $w$ is similar to the equation for the kinematic
pressure in the incompressible fluid, but now the compressibility depends on
the mechanism of the mass production in distributed sources. The evolution
of the density $\rho $ is determined by the continuity equation: $d\rho
/dt=-\alpha \rho +\mu $, where $\mu $ is the rate of mass production. One
natural possibility is that source is self-adjusting in such a way that $\mu
=\alpha \rho $ and fluid is incompressible. Let us stress that (2) is not
equation for ordinary fluid, but a model equation for distributed sources.
The physical nature of these sources can be different for different kind of
media. Some possible applications of distributed sources and sinks are
indicated below. Independently of an inducing mechanism of motion, if the
source is distributed, we can expect that various parts of it will move each
other. This simple interaction is reflected in equation (2). Dynamics of
distributed sources, potentially, may depend on other effects, including
deformation. Some generalizations of equation (2) are presented below.

The nonlinear partial differential equation (2), generally, is not easy to
solve, even numerically. The type of nonlinearity in (2) is the same as in
the equation for two-dimensional (2D) vorticity field $\omega =\Delta \psi $
in incompressible fluid ($\psi $ is the stream function). However, there is
a difference in the orientation of the components of velocity, been
expressed in terms of $\psi $ (instead of $\varphi $). We will see below
that this leads to essential difference in dynamics and, in some sense,
makes dynamics of 2D distributed sources more complicated than 2D vortex
dynamics. It is known that 2D vortex dynamics, generally, is nonintegrable
and chaotic$^{[1,4]}$. We can expect the same for 2D distributed sources in
general situation. To obtain some analytical solutions of (2), we will use
the fact that $\alpha $ is $LI$.

For 2D axisymmetrical flow we have: 
\begin{equation}
\alpha (t,r)=\frac 1r\frac{\partial (rv)}{\partial r}=\alpha _o(r_o),\frac{dr%
}{dt}=v  \tag{3}
\end{equation}
where $v=\partial \varphi /\partial r$ is the radial velocity, subscript ``$%
o $'' indicate initial value and in the second equation in (3) $r(t,r_o)$ is
the trajectory. Introducing Lagrangian variable $p(t,p_o)\equiv r^2(t,r_o)$, 
$p_o=r_o^2$, we rewrite (3) in the form: $\frac \partial {\partial p}(\frac{%
\partial p}{\partial t})=\alpha _o(r_o)\equiv \beta (p_o)$, where we used
that $d/dt\equiv \partial /\partial t\mid _{p_o}.$ Multiplication by $%
\partial p/\partial p_o$, gives: $\frac \partial {\partial p_o}(\frac{%
\partial p}{\partial t})=\frac \partial {\partial t}(\frac{\partial p}{%
\partial p_o})=\beta (p_o)\frac{\partial p}{\partial p_o}$. Solution of this
equation has the form: $\partial p/\partial p_o=\exp [\beta (p_o)t]$.
Integration of this solution with natural condition $p=0$ when $p_o=0,$and
time differentiation gives: 
\begin{equation}
p(t,p_o)=r^2(t,r_o)=\int_0^{p_o}\exp [\beta
(z)t]dz,2rv(t,r_o)=\int_0^{p_o}\beta (z)\exp [\beta (z)t]dz  \tag{4}
\end{equation}

Formulas (4) present Lagrangian description of the system. The same formulas
give Eulerian velocity $v(t,r)$ and $\alpha (t,r)=\alpha _o(r_o)$ in
parametric form ($r_o$ - parameter). We note that a differential rotation of
the system will not affect the obtained results. The solution depends on the
initial distribution $\alpha _o(r_o)$. For example, let $\alpha
_o(r_o)=\theta =const$ for $r_o\leq l$ and $\alpha _o(r_o)=0$ for $r_o>l$.
From (4) in the Lagrangian description we get $r(t,r_o)=r_o\exp (\theta t/2)$%
, $v(t,r_o)=(\theta r_o/2)\exp (\theta t/2)$ for $r_o\leq l$ and $%
r(t,r_o)=[l^2\exp (\theta t)+r_o^2-l^2]^{1/2}$, $v(t,r_o)=(\theta l^2/2)\exp
(\theta t)[l^2\exp (\theta t)+r_o^2-l^2]^{-1/2}$ for $r_o>l$. In the
Eulerian description $v(t,r)=\theta r/2$ and $\alpha (t,r)=\theta $ for $%
r\leq L(t)=l\exp (\theta t/2)$, for $r>L(t)$ we have $v(t,r)=(\theta
l^2/2r)\exp (\theta t)$ and $\alpha (t,r)=0$.

The main thing is that distributed source ($\theta >0$) is spreading, while
distributed sink ($\theta <0$) is shrinking. In contrast, for 2D
incompressible fluid, where the role of $LI$ plays the vorticity $\omega $,
axisymmetric flow is the stationary solution. Distributed 2D sinks and
sources potentially can model some convective systems in the atmosphere.
Particularly, such modeling can be useful for horizontally converging
updrafts, leading to concentration of vorticity (compare with Ref. 2 and
references therein), and for horizontally divergent downdrafts, which cause
aircraft accidents (see Ref. 5 and references therein).

For the tree-dimensional (3D) distributed sources with spherical symmetry,
instead of (3), we have: $\alpha (t,r)=\frac 1{r^2}\frac \partial {\partial
r}(r^2v)=\alpha _o(r_o),\frac{dr}{dt}=v$. Introducing $q(t,q_o)\equiv
r^3(t,r_o)$, $q_o=r_o^3$ and $\gamma (q_o)\equiv \alpha _o(r_o)$, we get
analogous solution:

\begin{equation}
q(t,q_o)=r^3(t,r_o)=\int_0^{q_o}\exp [\gamma
(z)t]dz,3r^2v(t,r_o)=\int_0^{q_o}\gamma (z)\exp [\gamma (z)t]dz  \tag{5}
\end{equation}

For the one-dimensional (1D) distributed sources we will get general
solution without any restriction of symmetry. First steps of obtaining
solution are the same as before for 2D and 3D: 
\begin{equation}
\alpha (t,x)=\frac{\partial v}{\partial x}=\alpha _o(x_o),\frac{dx}{dt}=v,%
\frac{\partial x}{\partial x_o}=\exp [\alpha _o(x_0)t]  \tag{6}
\end{equation}

At this step we can not assume that $x(t,x_o)=0$ when $x_o=0$, because,
generally, we do not have symmetry relative to the origin of the coordinate.
Instead, we have to impose a boundary condition in order to fix the velocity
of the system of observation. One possibility is condition: $v(t,\infty
)+v(t,-\infty )=0$. We will see that with symmetric initial distribution $%
\alpha _o(-x_o)=\alpha _o(x_o)$, this condition insures that $x(t,0)\equiv 0$%
. Time derivative of the last equation in (6) and integration over $x_o$
gives: 
\begin{equation}
v(t,x_o)=\int_0^{x_o}\alpha _o(z)\exp [\alpha _o(z)t]dz+u(t),u(t)\equiv
v(t,0)  \tag{7}
\end{equation}
Using indicated boundary condition, we have: 
\begin{equation}
u(t)=\frac 12\int_0^\infty \{\alpha _o(-z)\exp [\alpha _o(-z)t]-\alpha
_o(z)\exp [\alpha _o(z)t]\}dz  \tag{8}
\end{equation}
We see that $u(t)=0$ for symmetric initial distribution, as expected.
Integrating (7) with respect to time, we get: 
\begin{equation}
x(t,x_o)=x_o+\int_0^{x_o}\{\exp [\alpha _o(z)t]-1\}dz+\int_0^tu(\tau )d\tau 
\tag{9}
\end{equation}

Formulas (7) - (9) give general analytical description of the dynamics of
distributed sources. With the simplest symmetric initial distribution $%
\alpha _o(x_o)=\theta $ for $\mid x_o\mid \leq l$ and $\alpha _o(x_o)=0$ for 
$\mid x_o\mid >l$, we get solution analogous to the presented above for the
axisymmetric 2D flow. Description of interaction of several patches of
distributed sources and sinks will be presented elsewhere.

There is an analogy between distributed sinks and the gravitation, for which
we also have general 1D analytical solution in the Lagrangian variables$%
^{[6]}$. Relative accelerations of gravitational masses depend on distances
(in any dimension) similarly to relative velocities of sinks$^{[2,6]}$.
Homogeneous solutions of the general relativity with the cosmological
constant (CC)$^{[6]}$ behave similarly to (5) with $\gamma =const$.
Incorporation of the 3D distributed sources into cosmological modeling can
contribute to an explanation of the accelerated expansion of the Universe,
which is now a hot topic in physics (see Ref. 8 and references therein).

Let us consider a generalization of the equation for distributed sources: 
\begin{equation}
\frac{dJ_\sigma }{dt}\equiv \frac{\partial J_\sigma }{\partial t}+v\frac{%
\partial J_\sigma }{\partial x}=0,J_\sigma \equiv \frac{\partial v}{\partial
x}(\frac{\partial a}{\partial x})^\sigma =\alpha _o(a)  \tag{10}
\end{equation}
Here $a\equiv x_o(t,x)$ is the inverse trajectory and $\sigma $ is an
arbitrary parameter. With $\sigma =0$ we return to (6). With $\sigma =1$ we
get 1D analog of 3D vortex dynamics, for which $\omega _k\partial
a_i/\partial x_k$ is the $LI$. This $LI$ corresponds to the Kelvin's theorem
of conservation of the velocity circulation in the local (Cauchy) form$%
^{[8,9]}$. It potentially can lead to formation of singularities due to the
vortex stretching. With $\sigma =-1$ equation (10) gives the inertial
motion: $v(t,a)=v_o(a)$. From (10) we get equation for the deformation: $%
\frac \partial {\partial t}(\frac{\partial x}{\partial a})=\alpha _o(a)(%
\frac{\partial x}{\partial a})^{1+\sigma }$. Solution of this equation is: $%
\frac{\partial x}{\partial a}=[1-\sigma t\alpha _o(a)]^{-\frac 1\sigma }$.
Here we assume that $\sigma \neq 0$, the limit with $\sigma \rightarrow 0$
gives (6). From (10) we now have: 
\begin{equation}
\frac{\partial v}{\partial x}=\alpha _o(a)[1-\sigma t\alpha _o(a)]^{-1} 
\tag{11}
\end{equation}

We see that for $\sigma <0$, which includes inertial motion, we get
finite-time singularity with compression ($\alpha _o$ \TEXTsymbol{<}$0$).
For $\sigma >0$, which includes analog of 3D vortex dynamics, we get
finite-time singularity with stretching ($\alpha _o>0$). The intermediate
case (7) - (9) has both tendencies (shrinking sinks and spreading sources),
but avoids finite-time singularities. We note, that equation (10) can be
written in the form (1) with $f=\sigma \alpha ^2$ . Similar generalizations
can be made for 2D and 3D distributed sources and can be interpreted in
terms of the barotropic (or incompressible) fluid with corresponding
modification of $w$.

Finally, let us try one more step in generalization by using second order
spatial derivatives in $LI$: 
\begin{equation}
\frac d{dt}(\frac{\partial ^2v}{\partial x^2})=0,\frac d{dt}\equiv \frac
\partial {\partial t}+v\frac \partial {\partial x}  \tag{12}
\end{equation}
Assuming, for simplicity, boundary condition $v(t,-\infty )=0$ and
integrating (12) from $-\infty $ to $x$, we get: $\frac{d\alpha }{dt}=\frac
12\alpha ^2,\alpha \equiv \frac{\partial v}{\partial x}$. Solution of this
equation corresponds to (11) with $\sigma =1/2$. Surprisingly, we get
equivalence of two $LI$: $\partial ^2v/\partial x^2$ in (12) and $%
J_{1/2}=\partial v/\partial x(\partial a/\partial x)^{1/2}$in (10).

A more detailed analysis and interpretation of the obtained in this letter
nonlinear partial differential equations, as well as their generalization
with diffusion, will be presented elsewhere. Here we used them for
illustration of the Lagrangian modeling - from $LI$ to dynamics - and
corresponding technique. This approach creates a broad variety of Lagrangian
models for nonlinear continuous media.\vspace{0.2in}

\textbf{REFERENCES}\vspace{0.2in}

[1] E. A. Novikov, ``Stochastization and collapse of vortex systems'', Ann.
NY Acad. Sciences \textbf{357}, 47 (1980)

[2] E. A. Novikov and Yu. B. Sedov, ``Concentration of vorticity and helical
vortices'', Fluid Dynamics \textbf{18}, 6 (1983)

[3] A. E. Novikov and E. A. Novikov, ``Vortex-sink dynamics'', Physical
Review E \textbf{54}, 3681 (1996)

[4] E. A. Novikov and Yu. B. Sedov, ``Stochastic properties of a four-vortex
system'', Sov. Phys. JETP \textbf{48}(3), 440 (1978); ``Stochastization of
vortices'', JETP Lett. \textbf{29}(12), 678 (1979). S. L. Ziglin, ``The
nonintegrability of the problem on the motion of four vortices of finite
strengths'', Physica D \textbf{4}, 268 (1982). H. Aref, ``Integrable,
chaotic and turbulent vortex motion in two-dimensional flows'', Ann. Rev.
Fluid Mech. \textbf{15}, 345 (1983). E. A. Novikov, ``Chaotic vortex-body
interaction'', Phys. Lett. A \textbf{152}, 393 (1991). J. B. Kadtke, E. A.
Novikov, ``Chaotic capture of vortices by a moving body'', Chaos \textbf{3}%
(4), 543 (1993). E. A. Novikov, ``Vorticity gradient in 2D turbulence of
ideal fluid'', Phys. Lett. A \textbf{266}, 377 (2000)

[5] E. A. Novikov, ``Transformation of a vortex ring, initiated by a
downburst, into a horseshoe vortex in the boundary layer'', Boundary-Layer
Meteorol. \textbf{38}, 305 (1987)

[6] E. A. Novikov, ``Nonlinear evolution of disturbances in a
one-dimensional universe'', Sov. Phys. JETP \textbf{30}(3), 512 (1970)

[7] J. V. Narlikar, \textquotedblleft An Introduction to
Cosmology\textquotedblright\ (Cambridge University Press, Cambridge, 2002).
Einstein introduced CC to make the steady-state cosmology, but later (after
discovery of the expansion of the Universe) dismissed CC as his greatest
blunder. Careful observations of distant supernovae in the last five years
suggest the accelerated expansion of the Universe, which renewed interest to
CC and to other models with repulsion.

[8] A. Albrecht and C. Skordis, ``Phenomenology of realistic accelerating
universe using only Planck-scale physics'', Phys. Rev. Lett. \textbf{84},
2076 (2000); E. V. Linder, ``Exploring the expansion history of the
universe''$ibid.$ \textbf{90}, 091301 (2003); B. G. Sidharth, ``The new
cosmos'', Chaos, Solitons, Fractals \textbf{18}, 197 (2003); Y. Fujii and K.
Maeda, ``The Scalar-Tensor Theory of Gravitation'' (Cambridge University
Press, Cambridge, 2003).

[9] J. Serrin, ``Mathematical Principles of Classical Fluid Mechanics'', p.
152 [Handbuch Der Physik, Band 8(1), Ed. C. Truesdell; Springer, 1959]

[10] This $LI$ was used for the vorton method in 3D vortex dynamics [E. A.
Novikov, ``Generalized dynamics of three-dimensional vortical
singularities'', Sov. Phys. JETP \textbf{57}(3), 565 (1983); M. J. Aksman,
E. A. Novikov, S. Orszag, ``Vorton method in three-dimensional
hydrodynamics'', Phys. Rev. Lett. \textbf{54}, 2410 (1985)]. Statistics of
microcirculations was studied in E. A. Novikov, ``Microcirculations in
turbulent flows'', Phys. Rev. E \textbf{52}, 2574 (1995)

\end{document}